\documentclass[a4paper,11pt]{article}
\usepackage{pos}

\newcommand{\flc}{fluctuations }
\newcommand{\pt}{$p_{\rm T}$ }

\title{Mean transverse momentum fluctuations in pp collisions at $\sqrt{s}$ = 13 TeV using ALICE detector at LHC}
\ShortTitle{Mean transverse momentum fluctuations in pp collisions}

\author*[a]{Bushra Ali (for the ALICE Collaboration)}

\affiliation[a]{Aligarh Muslim University, Aligarh 202002, India}

\emailAdd{bushra.ali@cern.ch}

\abstract{
An analysis of the high-multiplicity triggered pp data at $\sqrt{s}$ = 13 TeV, obtained with the ALICE detector, is carried out to study the event-by-event fluctuations of mean transverse momentum ($\langle p_{\rm T}\rangle$) using a two-particle \pt correlator. The driving force behind these studies is the search for dynamical fluctuations that may be associated with the formation of Quark-Gluon Plasma (QGP) droplets in small collision systems, such as pp, for which traces have been observed in previous studies. The values of the correlator are observed to decrease with increasing charged particle density and exhibit power-law behavior similar to the one observed for pp and Pb--Pb collisions at lower energies. The findings also reveal that Monte Carlo model PYTHIA reproduces the observed dependence of the correlator on charged particle density.
 } 

\FullConference{%
  42nd International Conference on High Energy physics - ICHEP2024\\
  18-24 July, 2024\\
  Prague, Czech Republic
}
\begin{document}
\maketitle

\section{Introduction}
\label{intro}\vspace{-4mm}
Event-by-event mean transverse momentum \flc of charged particle produced in nucleus-nucleus (AA) collisions are regarded as a signal for a phase transition from a quark gluon plasma (QGP) to hadron gas~\cite{1,2}. Fluctuations of mean \pt have been observed to decrease with collision centrality. The observed trend of decrease, however, shows a deviation from what is expected from a simple superposition scenario. This has been attributed to the onset of thermalization and hardonization, but no strong connection to the critical behavior could be made. It has, however, been suggested that the \pt correlations and their centrality dependence might be affected by the initial state density fluctuations~\cite{1,3}. \\
\indent Fluctuations in $\langle p_{\rm T}\rangle$ might arise due to several kind of correlations among the \pt of the produced particles like resonance decay, quantum correlations and jets. To account for these correlations from conventional mechanisms, similar studies can be carried out for pp collisions where such correlations are also present~\cite{1}. The results from pp collisions can be used to construct a model independent baseline to search for the non trivial \flc in AA collisions. Furthermore, the evidence of deconfinement~\cite{4}, strangeness enhancement~\cite{5} and observation of collectivity in high-multiplicity pp collision events~\cite{6} has increased the interest in pp collision at LHC energies. An attempt is, therefore, made to study the mean  \pt \flc in high-multiplicity events in pp collisions at $\sqrt{s}$ = 13 TeV using the two-particle correlator, $C_m$, which is regarded as a measure of dynamical component of mean \pt fluctuations. The study is based on the analysis of 2 billion minimum-bias and 0.8 billion high-multiplicity events recorded by ALICE detector. \\
\indent In a multiplicity class m, $C_{m}$, is defined as~\cite{1}
\vspace{-2mm}
\begin{equation}
C_m = \frac{1}{\Sigma_{k}^{n_{ev},m} N_k^{pairs}} \Sigma_{k=1}^{n_{ev},m}\Sigma_{i=1}^{N_{acc},k}\Sigma_{j=i+1}^{N_{acc},k} \left(\left(p_{\rm T, i} - M(p_{\rm T})_m\right) . \left(p_{\rm T, j} - M(p_{\rm T})_m\right)  \right)
\end{equation}
where, $N_k^{pairs}$ is the number of pairs in an event, $n_{ev}$ is the number of events in class $m$ and $M(p_{\rm T})_m$ is the mean $p_{\rm T}$ of all particles in all events of that multiplicity class and is given as
\vspace{-2mm}
\begin{equation}
M(p_{\rm T})_m = \frac{1}{\Sigma_{k=1}^{n_{ev},m} N_{acc, k}} \Sigma_{k=1}^{n_{ev},m}\Sigma_{i=1}^{N_{acc},k} p_{\rm T,i}
\end{equation}
where, $N_{acc,k}$ is the measured multiplicity of an event $k$.
\vspace{-5mm}
\section{Results and Discussion}
\vspace{-4mm}
\begin{figure*}[ht]
    \centering
        \includegraphics[width=0.44\linewidth,clip]{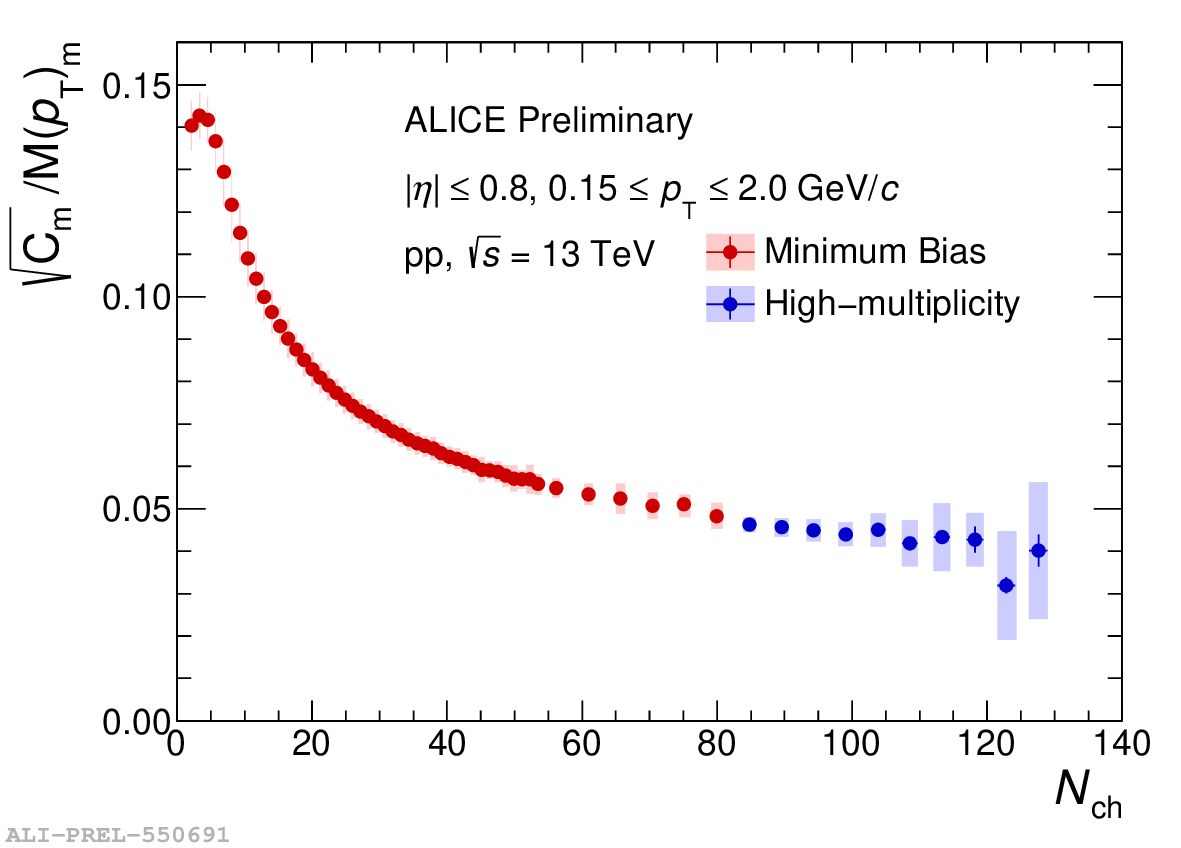}
        \includegraphics[width=0.44\linewidth,clip]{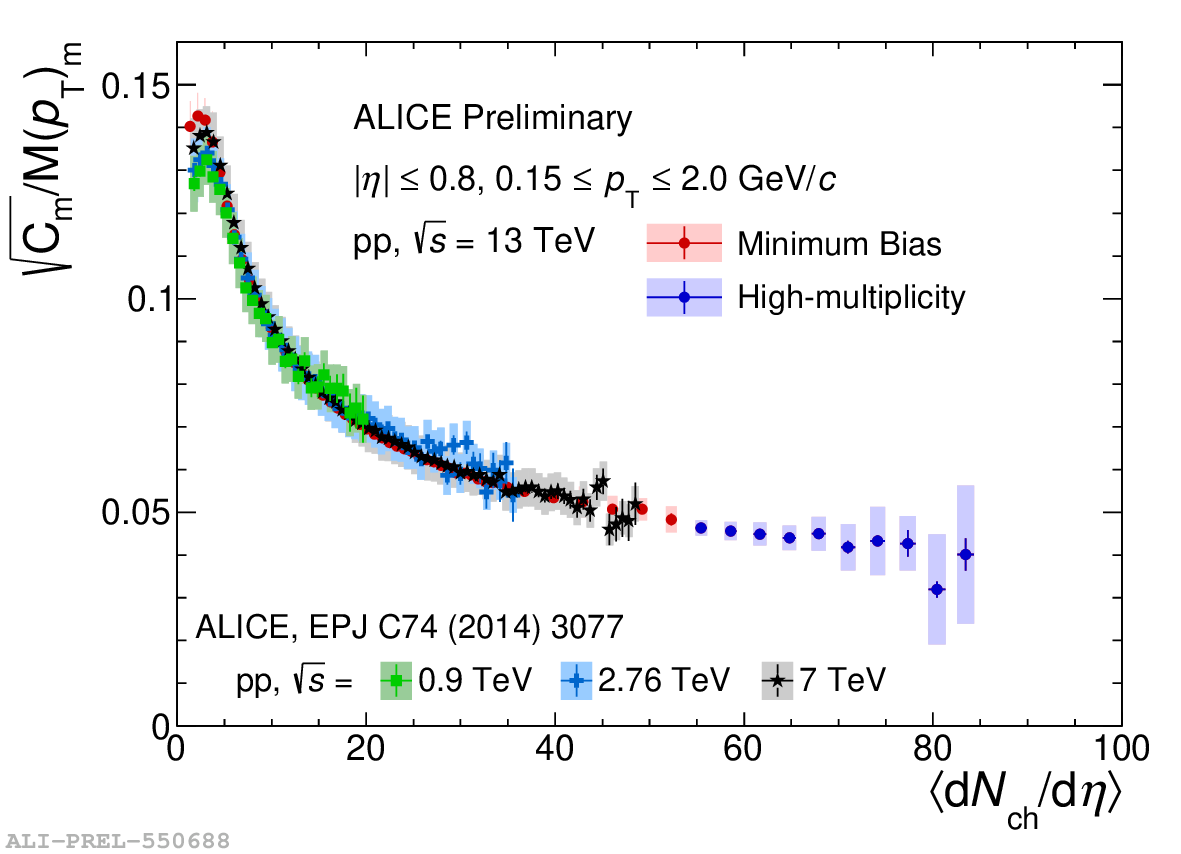}
        \caption{{\footnotesize  Variations of correlator with multiplicity for pp collisions at $\sqrt{s}$ = 13 TeV (left). A comparison of correlator values for pp collisions at various energies. The statistical errors are represented by lines and systematic errors by shaded boxes.}}\label{fig:r1}
\end{figure*}
\indent Variations of the normalized correlator, $\sqrt{C_m}/M(p_{\rm T})_m$ on the true multiplicity for minimum bias (MB) and high-multiplicity (HM) triggered events for 13 TeV pp collisions are shown in left panel of Fig.~\ref{fig:r1}. The measurement has been performed in intervals of one unit of $N_{\rm ch}$ up to $N_{\rm ch}$ = 54 and beyond this value, in intervals of four units to minimize statistical fluctuations in the high-multiplicity tail. It is interesting to note that the MB and HM data points tend to form a continuous curve and give a common trend of decrease of correlator with increasing multiplicity. The non-zero values of the correlator is an indication of presence of significant dynamical fluctuations in the data. The observed dependence of $\sqrt{C_m}/M(p_{\rm T})_m$ on the mean charged particle density, d$N_{\rm ch}$/d$\eta$ is compared with those reported at lower energies~\cite{2}. The results are shown in the right panel of Fig.~\ref{fig:r1}. The figure shows that the correlator acquires nearly energy independent values except in the region of low multiplicity ($\langle$d$N_{\rm ch}$/d$\eta\rangle \leqslant$ 8), where the difference is $\sim10\%$, whereas for $\langle$d$N_{\rm ch}$/d$\eta\rangle\geqslant$ 10 the difference may be noted to be $\sim2\%$ only.\\
\begin{figure*}[h]
    \centering
        \includegraphics[width=0.44\linewidth,clip]{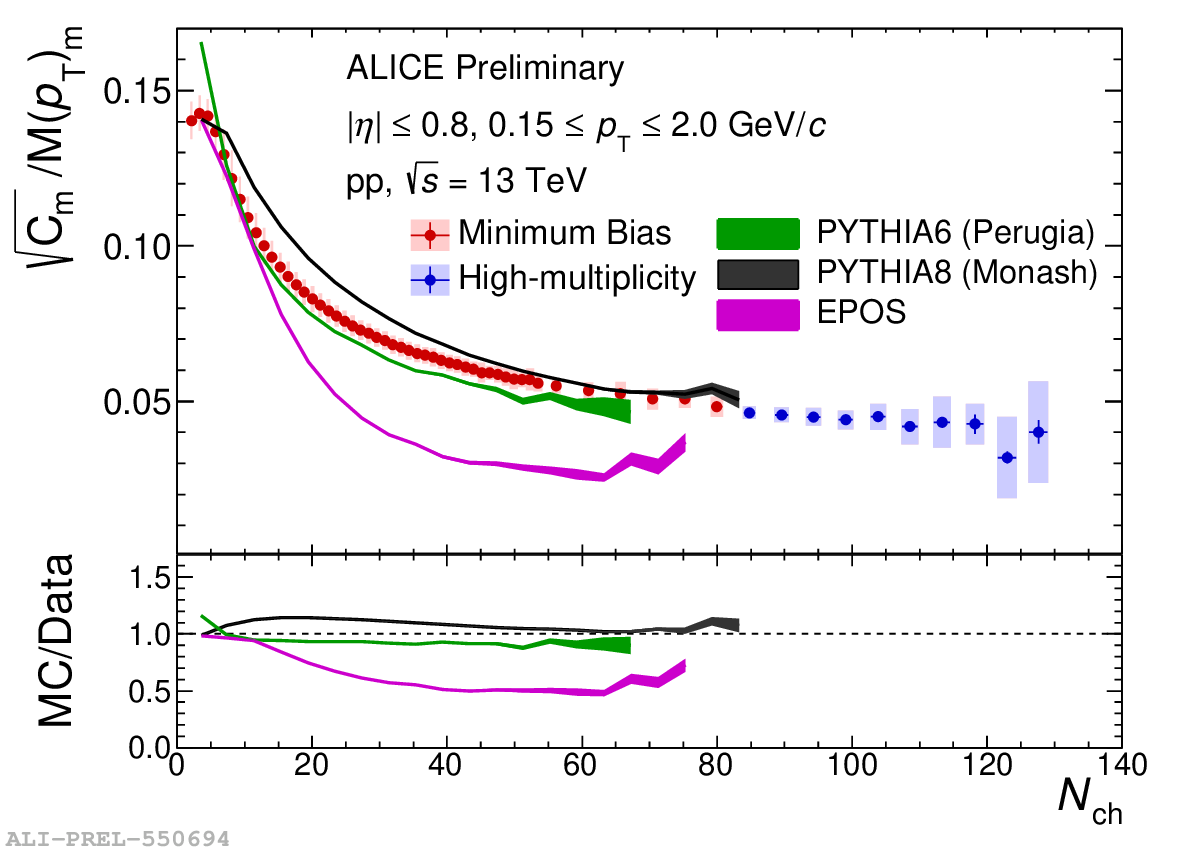}
        \includegraphics[width=0.44\linewidth,clip]{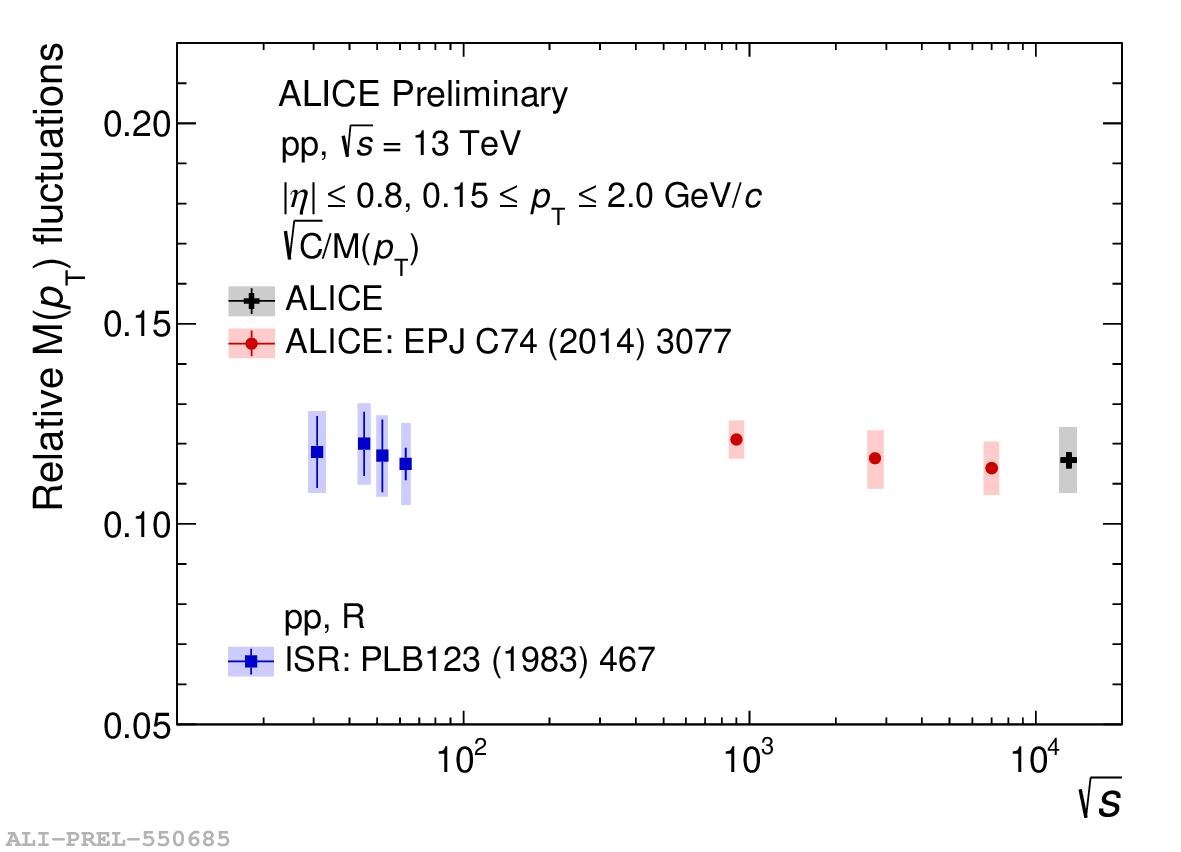}
        \caption{{\footnotesize Models predictions compared with the experimental findings (left). Inclusive correlator values from ISR to LHC energies (right). The statistical errors are represented by lines and systematic errors by shaded boxes.}}\label{fig:r2}
\end{figure*}
\indent Shown in Fig.~\ref{fig:r2} (left) is the comparison of the present findings with the predictions of Monte Carlo (MC) models PYTHIA (Monash and Perugia tunes) and EPOS. It is observed that for $N_{\rm ch} \leq $ 15, EPOS satisfactorily reproduces the data but predicts quite small values for $N_{\rm ch} > $ 18 and beyond. On the other hand, PYTHIA8 describes the data qualitatively while PYTHIA6 predictions agree both qualitatively and quantitatively. Inclusive values of the correlator are also computed and compared to those reported for lower energies. This is displayed in right panel of Fig.~\ref{fig:r2}. The blue markers in the figures are the values of relative fluctuation measure, R reported at ISR energies~\cite{7}; it has been shown~\cite{1} that \flc estimated using R and $\sqrt{C_m}/M(p_{\rm T})_m$ agree with each other within $10-15\%$. It is evidently clear from the figure that the correlator acquire almost energy independent values from ISR to LHC energies.
\end{document}